\documentclass[aps,amsfonts,amsmath,tightenlines,twocolumn,showpacs,nofootinbib]{revtex4}
\usepackage{epsf}

\def\be{\begin{equation}}
\def\ee{\end{equation}}
\def\beq{\begin{equation}}
\def\eeq{\end{equation}}
\def\bea{\begin{eqnarray}}
\def\eea{\end{eqnarray}}
\def\bml{\begin{subequations}}
\def\blea{\bml\begin{eqnarray}}
\def\elea{\end{eqnarray}\end{subequations}}

\begin{document}

\title{Numerical search for a fundamental theory}

\author{Vitaly Vanchurin}
\email{vitaly@cosmos.phy.tufts.edu}

\affiliation{Arnold-Sommerfeld-Center for Theoretical Physics, Department f\"ur Physik, 
Ludwig-Maximilians-Universit\"at M\"unchen, Theresienstr. 37, D-80333, Munich, Germany}

\begin{abstract}

We propose a numerical test of fundamental physics based on  the complexity measure of a general set of functions, which is directly related to the Kolmogorov (or algorithmic) complexity studied in mathematics and computer science. The analysis can be carried out for any scientific experiment and might lead to a better understanding of the underlying theory. From a cosmological perspective, the anthropic description of fundamental constants can be explicitly tested by our procedure. We perform a simple numerical search by analyzing two fundamental constants: the weak coupling constant and the Weinberg angle, and find that their values are rather atypical.
 
\end{abstract}


\maketitle 

\section{Introduction}

The search for a fundamental theory of nature has puzzled many generations of scientists.  Over time, mathematics describing physics became extremely complicated, but nevertheless very successful. Due to complexity, many branches of physics are evolving without much overlap simply because it is hard for a single human being to keep track of all of the new developments. Unfortunately, that might also mean that we could be missing something very fundamental that relates different fields, different theories, or even different constants of nature in a somewhat trivial way.

In this article we intend to test if the constants of nature are truly independent. In Ref. \cite{Tegmark} an overall twenty six fundamental constants are listed, with presumably another six required to fit the upcoming data. A natural question to ask is whether all of these constants are really independent, or if some of them can be expressed in terms of the others.  Ideally, one would hope to reduce the number of  fundamental constants to just a few, the fewer, the better.

If there are no fundamental reasons that fix some constant $C$ to the measured value $C_m$, then we should be able to vary it and still obtain corresponding results. In other words, if we leave all of the physics unchanged, and add $\Delta$ to $C_m$, then the modified theory, with the constant given by $C=C_m+\Delta$, should approach the original one in the limit $\Delta \rightarrow 0$. 

On the other hand,  if the observed value of the constant $C_m$ is described by an anthropic selection process \cite{Weinberg, Vilenkin}, then it must be obtained from some probability distribution $P(C)$. If $P(C)$ is smooth, then in a small enough neighborhood of $C_m$ the function is nearly flat. This means that we are equally likely to observe an arbitrary value of $C$ within a small neighborhood of the measured value $C_m$. Thus, the anthropic and non-anthropic views of fundamental physics seem to support the following hypothesis:
A fundamental constant $C$ must not lie in a preferred location on the interval $(C_m-\Delta, C_m+\Delta)$ for small $\Delta$.

In the following sections, we will define exactly what we mean by preferred location. In contrast, if we find that some constant takes a highly atypical value within its neighborhood, then the constant in question may not be fundamental, or its anthropic function $P(C)$ is highly irregular.

The paper is organized as follows. In the second section we describe the basic idea by preforming a simple numerical search. In the third section we extend the analysis to an arbitrary pair of fundamental constants and apply the procedure to a simple model. In the forth section a general mathematical formalism is developed. In the last section we summarize the main results.

\section{Motivations}

Measurements in any experiment could be described by a set of real numbers with some uncertainties\footnote{All physical quantities are given in Planck units with $c=\hbar=G=1$.}. If some experiment is explained by a theory, then there are measurements that are related by expressions derived from the underlying theory. On the other hand, there are always some other measurements that are completely independent of one another, based on the theory in question.  We will refer to such measurements as fundamental constants of the theory. 

\subsection{Numerical example}

To quickly motivate the analysis of this paper we consider two fundamental constants: the weak coupling constant $g$ at $m_Z$ and the Weinberg angle $\theta_W$. In dimensionless units the constants take the following values \cite{Tegmark}:
\be
\begin{aligned}
& g = 0.6520 \pm 0.0001 \\
& \theta_W = 0.48290 \pm 0.00005
\end{aligned}
\ee
We would like to perform a numerical test of the precision intervals $(0.6519, 06521)$ and $(0.48285, 0.48295)$, to check weather these intervals are typical or weather they are ``carefully chosen". (Intuitively, the intervals $(3.1415926, 3.1415928)$ or $(0.9999999, 1.0000001)$ seem to be very atypical, whereas the interval $(4.2831741, 4.2831743)$ is in some sense more random.)

Let us choose only three constants ${\pi, e, 1}$ and five elementary functions:
\be
\begin{aligned}
& f_1(x,y)=x+y, \\
& f_2(x,y)=x-y, \\
& f_3(x,y)=x y, \\
& f_4(x,y)=x/y, \\
& f_5(x)=\sqrt{x}.
\label{functions}
\end{aligned}
\ee
For the time being we define the complexity measure $M$ of expressions as the total number of elementary functions in the expression. For example,
\be
\begin{aligned}
&M[\frac{2+\pi}{e^2-1}]=M[\frac{(1+1)+\pi}{(e \cdot e) - 1}]=\\
&M[f_4(f_1(f_1(1,1),\pi), f_2(f_3(e,e),1))]=5.
\end{aligned}
\ee
The numerical complexity of an interval is then defined as the smallest complexity of all expressions that generate a number in the interval. For example, the complexity of the interval  $(0.9999999, 1.0000001)$ is $0$, and the complexity of $(1.9999999, 2.0000001)$ is $M[f_1(1,1)]=1$.

\begin{figure}
\begin{center}
\leavevmode\epsfxsize=4.0in\epsfbox{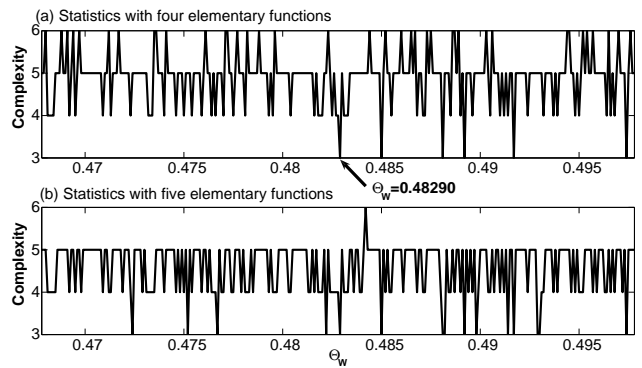}
\end{center}
\caption{Numerical complexities of 300 intervals around the measured value of  the Weinberg angle.}
\label{fig:w}
\end{figure}

We have developed a computer program which can search through all possible expressions with complexity up to $6$. The numerical complexities of 300 neighboring intervals around the measured value of $\theta_W$ are plotted on Fig.\ \ref{fig:w}
. All five elementary functions of Eq.~(\ref{functions}) are used to generate the plot on Fig.\ \ref{fig:w}b, and only the first four functions are considered for the plot on Fig.\ \ref{fig:w}a. In both cases, it appears that the numerical complexity of the interval $(0.48285, 0.48295)$ is way too small, when compared to other intervals: ... , $(0.48265, 0.48275)$, $(0.48275, 0.48285)$, $(0.48295, 0.48305)$, $(0.48305, 0.48315)$, ... . In fact, the Weinberg angle is given by a simple expression:
\be
\theta_W = \frac{2}{1+\pi}.
\ee
with complexity measure equal to 3. 

The probability of observing such a small value of the complexity in the neighborhood of $\theta_W$ is about $1.7\%$ for the analysis with four elementary functions. (It could be seen directly from the Fig.\ \ref{fig:w}a: only five out of 300 neighboring intervals have the complexity equal to 3.) The probability of observing the complexity of $(0.48285, 0.48295)$ in the five functions analysis of Fig.\ \ref{fig:w}b is a bit higher and equal to $4.7\%$. Of course, that could have happened by chance, and thus one needs a better measurement of the Weinberg angle in order to prove, or disprove the above relation.

\begin{figure}
\begin{center}
\leavevmode\epsfxsize=4.0in\epsfbox{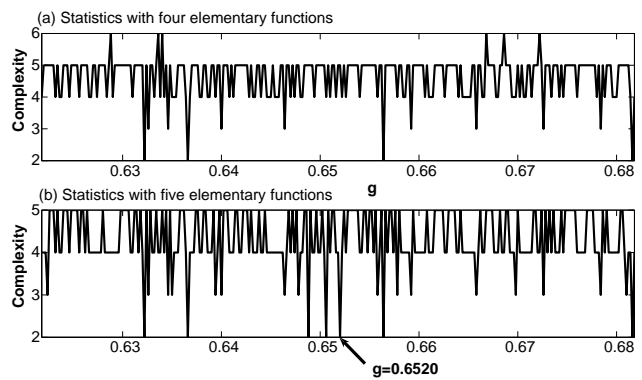}
\end{center}
\caption{Numerical complexities of 300 intervals around the measured value of  the weak coupling constant at $m_Z$.}
\label{fig:g}
\end{figure}

The numerical complexity of the precision interval of $g$ seems to be very typical in its neighborhood from the analysis with four functions (Fig.\ \ref{fig:g}
a), and very atypical (the probability is $2\%$) from the analysis with five elementary functions (Fig.\ \ref{fig:g}b). The following simple expression was revealed:
\be
g = \frac{\sqrt{\pi}}{e}.
\ee
Once again, in order to prove or disprove the relation a much more precise measurement of $g$ must be available.

\subsection{Discussion} 

The reader must be wondering if we really propose that $g$ and $\theta_W$ are given by our simple expressions. The short answer is no. We have chosen the weak coupling constant and the Weinberg angle only because out of all constants listed in Ref. \cite{Tegmark}, they are measured with the highest precision. Nevertheless, the uncertainty was too large to draw a definite conclusion. Presumably, with better technology we can hope to get much better measurements of the constants. It should also be noted that the precision could be dramatically improved by simply repeating the experiment. This is true even if the uncertainty principle of quantum mechanics does not immediately allow to measure some constants precisely. If the precision of a single measurement is $\delta$, the combined uncertainty of $N$ measurements is given by $\delta/\sqrt{N}$, and thus can be arbitrary small.

Unfortunately, we were not able to study the properties of the fine structure constant, which is already measured with very high precision: $\alpha=0.007297352568 (24)$ \cite{alpha}. Not even a single expression with $M<7$ was found to generate a number inside of the precision interval of $\alpha$. To calculate the relation function of such a tiny interval, in general, the functions $f \in \cal{G}$ with much larger complexities must be checked, which is currently out of reach. Perhaps more advanced algorithms, such as PSLQ \cite{PSLQ}, should be used to speed up the numerical search.

Similar parameter fitting was first performed in Ref. \cite{Eddington}, where the expression $\alpha=1/136$ was claimed. The major difference in here, is that the statistical significance of such claims is actually verified. In fact, the "discovery" of Ref. \cite{Eddington} could have been easily checked for a given set of elementary functions. Fortunately, the expression $\alpha=1/136$ (and  $\alpha=1/137$) is already ruled out and we do not have to perform the analysis. However, there are other numerical coincidences that have been discussed in the literature. For example, in Ref. \cite{Noyes}, the Weinberg angle is given by the expression $sin(\theta_W) = \frac{1}{2}-\frac{1}{42}$ which has a relatively large complexity. This should not be regarded as atypical, when the expressions with much smaller complexities could also describe the constant in question. 

\section{Relation functions}

In the previous section we have analyzed the numerical structure of some of the fundamental constants through the complexity of  their precision intervals. The proposed procedure enabled us to search for the numerical expressions that  could explain the observed values of these constants. In addition, we were able to quantify the statistical significance of the obtained simple expressions. In this section we will generalize the approach to study the numerical structure of an arbitrary pair of fundamental constants.

\subsection{Two intervals example}

For simplicity, let us consider two fundamental constants given by positive real numbers $C_1$ and $C_2$ with precisions $\delta_1/2=\delta_2/2=\delta/2$. We assume that $\delta \ll |C_1| <  |C_2|$ or, in other words,  $C_1$ and $C_2$ are measured very precisely. In what follows, we will discuss a simple procedure which could give a definite answer to the question of whether constants $C_1$ and $C_2$ are really independent.

We define a simple set of real valued functions and a measure defined on them. Let $\cal{F}$ be the set of functions of the form $f(x)=m x$ and their inverses  $f^{-1}(x)=x/m$, where $m \in \mathbb{N}$. The measure $M$ is a map from $\cal{F}$ to $\mathbb{N}$, defined as $M(f)=m$ if $f(x)=m x$ or $f(x)=x/m$.

We can now define a two-interval ($I_1$ and $I_2$) relation\footnote{Unlike the correlation function, which is defined on a set of points, the relation function is defined on a set of intervals.} function $R(I_1,I_2)$ with respect to the set of functions $\cal{F}$ and measure $M$:
\be
\begin{aligned}
R(I_1, I_2)= \min\{&M(f_1)+M(f_2) | \\
& \exists z \in \mathbb{R}: f_1(z)\in I_1\land f_2(z)\in I_2\}. 
\end{aligned}
\ee
(In other words, we are trying to approximate the ratio $C_1/C_2$ with two integers $m_1/m_2 \approx C_1/C_2$, such that $m_1+m_2$ is the smallest.)

\begin{figure}
\begin{center}
\leavevmode\epsfxsize=3.5in\epsfbox{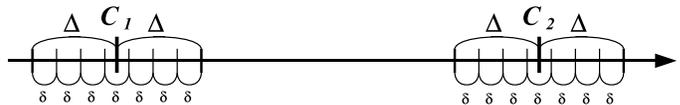}
\end{center}
\caption{Expectation value of the two-interval relation function is calculated by averaging over all pairs of $\delta$ intervals inside $\Delta$-neighborhoods of $C_1$ and $C_2$.}
\label{fig:example}
\end{figure}

In what follows, the two-interval relation function of the uncertainty intervals of $C_1$ and $C_2$ is denoted as:
\be
R_0 \equiv R[(C_1-\frac{\delta}{2},C_1+\frac{\delta}{2}),(C_2-\frac{\delta}{2},C_2+\frac{\delta}{2})].
\ee 
To calculate an expectation value of $R_0$, we have to average the relation function over all pairs of $\delta$ intervals inside larger neighborhoods ($\Delta \gg \delta$) of $C_1$ and $C_2$ (See Fig.\ \ref{fig:example}
). In general, a numerical search must be performed in order to estimate the expectation value, but for our simple example we can find it analytically. 

The aim is to find two integers $m_1$ and $m_2$ corresponding to functions $f_1(x)=m_1 x$ and $f_2(x)=m_2 x$, such that for some $x \in \mathbb{R}$:
\be
\begin{cases}
\;|C_1-x m_1| < \frac{\delta}{2} \\
\;|C_2-x m_2| < \frac{\delta}{2} \\
\end{cases}
\ee
In order to minimize $M(f_1)+M(f_2)=m_1+m_2$, it is enough to find the smallest $m_2$, for which both inequalities are satisfied (for some $m_1$). Since $C_2>C_1$, then $m_2 > m_1$, and the two inequalities can be reduced to a single one:
\be
| C_1 - \frac{C_2}{m_2}m_1 | < \delta.
\label{eq1}
\ee

For any fixed $m_2$, the probability that the inequality in Eq.~(\ref{eq1}) is satisfied (for some $m_1$) can be written as
\be
 P(m_2)=\frac{m_2 \delta}{C_2}.
\ee
Thus, one would expect the inequality to hold for at least one value of $m_2$ if
\be
\sum_{k=1}^{m_2} \frac{k \delta}{C_2} \gtrsim 1
\ee
or
\be
m_2 \gtrsim \sqrt{\frac{C_2}{\delta}}.
\ee
This immediately implies that the expectation value of the relation function is approximately given by:
\be
E \equiv m_2 + m_1 \sim m_2 + m_2 \frac{C_1}{C_2} \sim \frac{C_1+C_2}{\sqrt{\delta C_2}}.
\label{eq2}
\ee

For a particular choice of $C_1$, $C_2$ and $\delta$ there are three possible outcomes of the procedure: $R_0 \sim E$,  $R_0 \ll E$,  $E \ll R_0$, which we will discuss separately.

\subsection{Interpretations}

Naturally, one would expect $R_0 \sim E$, if constants $C_1$ and $C_2$ are not simply related with respect to the set of functions $\cal{F}$ and measure $M$.  However, even then it is still possible that $C_1$ and $C_2$ are related by some other set of functions with respect to some other measure. The procedure for generating the most general set of functions is rather complicated and will be discussed in the following section. 

By far, the most important possible outcome of our procedure is the discovery that $R_0 \ll E$. In this case, it is clear that we have reached a contradiction and something is completely wrong with our understanding of the theory. The numbers that we thought are not related are in fact related by some very simple relation. After this discovery, one can go ahead and study the functional relation between $C_1$ and $C_2$, in order to reformulate the theory using this hint. (For example, Mendeleev's periodic table of elements could have been discovered by applying the discussed procedure with exactly the complexity measure on the set of functions that we have introduced in this section \cite{Mendeleev}. The atomic masses of elements are simply related to each other with respect to  $\cal{F}$ and measure $M$, and thus  $R_0 \ll E$.)

Another interesting result of the procedure is somewhat more controversial.  $E \ll R_0$ implies that $C_1$ and $C_2$ are atypically unrelated (or anti-related). It means that they are ``placed''  in specific locations on the real line such that one cannot simply relate them by a given set of functions. This might be a direct evidence that our set of functions and/or the measure procedure is not suited to study the particular pair of constants.  Alternatively, it might also mean that one has to introduce yet another (more) fundamental constant, that would make measured values of  $C_1$ and $C_2$ more typical in their neighborhoods.

\section{General analysis}

Let us denote the general set of functions by $\cal{G}$. Every element of the set $f$ is a function constructed from a finite combination of its fundamental members: numbers (a finite number of numbers), functions (elementary operations, exponential function and special functions), loops (unevaluated integrals, sums and products),  recursive relation and inverses of all of the above. Note that the set of fundamental elements is finite, when the set of general functions is infinite, but countable. Each function $f$ of the set  $\cal{G}$ can be represented as a binary tree with functions associated with nodes (one or few) of the tree. The complexity $M$ of such tree can, for example, be defined as the sum of complexities of all nodes.  

With respect to $\cal{G}$ and some complexity measure $M$ we can construct a two-interval relation function $R$:
\be
\begin{aligned}
&R(I_1,I_2)=\min\{M(f)| \exists z_1, z_2 \in \mathbb{R}:\\
&|z_1-C_1|<\frac{\delta_1}{2} \land |z_2-C_2|<\frac{\delta_2}{2} \land f(z_1,z_2)=0\}.
\end{aligned}
\ee
Similarly, we can generate an $n$-interval relation function:
\be
\begin{aligned}
&R(I_1,..., I_n)=\min\{M(f)|\;\exists z_1...z_n \in \mathbb{R}:\\
&|z_1-C_1| < \frac{\delta_1}{2} ... |z_n-C_n| < \frac{\delta_n}{2} \land f(z_1...z_n)=0 \}, 
\end{aligned}
\ee
where $n\in\mathbb{N}$. Altogether, there are $2^n -1$ independent relation functions with respect to the precision intervals. If all $n$ constants are really independent, then all of the relation functions $R_1, R_2, ... R_{2^n -1}$ of uncertainty intervals $(x_i-\frac{\delta_i}{2}, x_i+\frac{\delta_i}{2})$ should be of the same order as their respective expectation values $E_1, E_2, ... E_{2^n -1}$. These values are calculated by averaging over all intervals:
\be
\begin{aligned}
I_i(d_i) \equiv (x_i+(d_i-\frac{1}{2})&\delta_i, x_i+(d_i+\frac{1}{2})\delta_i) \\
& \subset (x_i-\Delta_i, x_i+\Delta_i),
\end{aligned}
\ee
where $d_i\in\mathbb{Z}$. For example, the expectation value of the relation function for all $n$ constants is given by:
\be
E_{2^n -1}\equiv \frac{1}{N}\sum_{d_1}...\sum_{d_n } R(I_1(d_1),..., I_n (d_n)),
\ee
where $N$ is the normalization constant.

In order to compare the values of relation functions, intervals of the same size must be used. The relation function is larger for a smaller interval, as we have seen in the previous sections (Eq.~(\ref{eq2})). If we take the size of an interval to zero, the relation function either converges to a finite number, or diverges to infinity. In the former case the respective constants are proved to be related, whereas in the latter case the constants are proved to be independent. Of course  physically it is not possible to perform a measurement with infinite precision, and therefore we can only prove the dependence or independence to a certain degree.

To study the structure of a set of constants we might need to know more than just expectation values of the relation functions. It is straightforward to define a probability distribution $P_k$ of relation functions in $\Delta_i$ neighborhoods of the measured values of constants $C_i$'s.  For a given probability distribution $P_k(R)$, we can calculate the probability of observing a measured value of $R_k$ or smaller:
\be
S_k=\int_{0}^{R_k}P_k(R) dR.
\ee
If $S_k \ll 1$ or  $1-S_k \ll 1$, then the constants in question are atypically related or anti-related respectively. However, we have to be very careful, because for a large number of relation functions a very atypical situation may occur according to the central limit theorem. 

In order to draw a definite conclusion, we have to average separately the values of $S_k$ in the ranges  $S<\frac{1}{2}$ and  $S>\frac{1}{2}$:
\be
{\cal R}=\frac{\sum_{k=1}^{2^n -1} S_k\Theta(\frac{1}{2}-S_k)}{\sum_{k=1}^{2^n -1}\Theta(\frac{1}{2}-S_k)}
\ee
\be
{\cal A}=\frac{\sum_{k=1}^{2^n -1} (1-S_k)\Theta(S_k-\frac{1}{2})}{\sum_{k=1}^{2^n -1}\Theta(S_k-\frac{1}{2})}
\ee
where $\Theta(x)$ is a step function:
\be
\Theta(x)=  \begin{cases} & 0 \;\;\;\;\;\;\;\;\text{if}\;\; x \leq 0 \\
& 1 \;\;\;\;\;\;\;\; \text{if}\;\; x>0.
\end{cases}
\ee
Strictly speaking, ${\cal R}$  and  ${\cal A}$ are only defined if the denominators are non-zero. This is the case for a large number of fundamental constants, and that is exactly when these expressions are particularly useful. We call ${\cal R}$ and ${\cal A}$ the relation and anti-relation indexes of a given set of constants. 

Some interpretations of different outcomes of our procedure on a set of fundamental constants were given above. We can summarize the possibilities in the language of relation and anti-relation indexes. If the constants are: \\
	- independent, then ${\cal R} \sim {\cal A} \sim 1$,\\
	- related, then ${\cal R} \ll 1$ and ${\cal A}\sim 1$,\\
	- anti-related, then ${\cal R}\sim 1$ and ${\cal A}\ll 1$,\\
	- related and anti-related, then ${\cal R}\ll 1$ and ${\cal A}\ll 1$.

In the above analysis we have assumed that all relation functions are uncorrelated. If correlations between some relation functions are not negligible, the definitions for relation and anti-relation indexes have to be adjusted.

\section{Conclusion}

In this article we have developed a procedure of testing whether the fundamental constants are really fundamental. The method relies on the hypothesis that the measured value of a fundamental constant must not lie in a preferred location within its neighborhood. If the hypothesis is proved to be wrong, then the anthropic function (if exists) must be sharply peaked around the measured value of the constant. Similarly, a non-anthropic point of view implies that some constants of nature are very carefully chosen. Both approaches suggest that one might be able to derive the values of fundamental constants from first principles.

As we have discussed above, our analysis relies on the definition of the complexity which is not unique.  Nevertheless, there are statements that one can make regardless of the exact definition, given that all of the finite expressions have a finite complexity. In the limit when the size of the interval goes to zero, its complexity grows to infinity. This is true for all of the intervals, unless the interval must contain (due to some fundamental reason) a constant which is given by some fundamental expression. If the expression exists, then for very precise measurements of the fundamental constants it is guaranteed to be discovered. (In this case, an expectation value of the relation function can grow without bounds, whereas the relation function of the intervals in question is bounded from above.) 

Although we have shown that in principle the exact definition of the complexity is not crucial, in reality the complexity must be defined such that a useful result is obtained as soon as possible. Thus, it is important to carefully choose the complexity measure, which would unable us to tackle the problem efficiently. Fortunately, similar ideas had been already proposed in mathematics and computer science \cite{Kolmogorov, Solomonoff, Chaitin}. The algorithmic (or Kolmogorov) complexity is defined as a measure of the computational resource (e.g. the length of the shortest program) needed to specify an object. The relation of the complexity of intervals to the Kolmogorov complexity is an important question to be discussed in the upcoming publication.

The last point that we would like to address is not completely settled. The question is whether in practice it is possible to use our procedure to check widely accepted theories and/or to discover new physics. The discussed procedure relies very strongly on two parameters: precision of measurements and computational resources. One has to perform very precise measurements of fundamental constants and then run the computer program until a meaningful result is produced.  Both of these parameters depend very strongly on the state of available technological resources. However, even if we will discover that it is not possible to perform our analysis at present, this will undoubtedly change in the future, perhaps with help of quantum computers.

\section*{Acknowledgments}

It is a pleasure to thank Philipp Loewenfeld, Ken Olum, Max Tegmark and Serge Winitzki for stimulating discussions, and Alex Sousa for helpful comments. This work was supported in part by project ``Transregio (Dark Universe)''.

\end{document}